# Stratification and segregation under laminar convection

**A I Fedyushkin**

Ishlinsky Institute for Problems in Mechanics of the Russian Academy of Sciences, Moscow 119526, Russia

E-mail: fai@ipmnet.ru

**Abstract**. The paper is devoted to the study of the formation of stratification in an incompressible fluid due to convective laminar flows in horizontal layers heated from the side. Medium and intensive modes of stationary laminar thermal, concentrational and thermo-concentrational (in particular thermohaline) convection are considered, in which nonlinear flow features are manifested, which can radically change the flow structure and the characteristics of heat and mass transfer. The solutions of the problems of laminar problems of convection show the features of the formation of layered structures, vertical temperature and concentration stratification depending on the determining dimensionless parameters. The metastable instability of the direction of the stratification vector (the location of the free surface) in weightlessness in the presence of capillary convection was shown.

## 1. Introduction

In an initially homogeneous liquid located in the gravity field when heat or mass is supplied, vertical stratification in density may occur due to convective mixing. Vertical inhomogeneity in horizontal liquid layers can be induced both by only thermal (or concentration) convection and by the combined action of heat and impurity concentration and depends on the properties of the liquid and the intensity of convection [1–5]. In the case of thermo-concentration convection, in addition to the values of heat and mass flows, an important factor is the direction of the gradients of heat and concentration relative to the gravity vector. In [4], the variety of convective regimes is shown, even in the two-dimensional case, depending on the direction of heat and mass flows, and their classification is also given. Possible mechanisms of thermal, concentration and thermo-concentration convection in a thermostat during crystallization of calcium phosphates from a solution with the formation of layered convective structures are shown in paper [5].

For many hydrodynamic processes with convective heat and mass transfer, an important aspect is the occurrence of temperature or concentration stratification and the formation of stratified flows in the volume of liquid [1]. There is a great variety of convective stratified currents in nature that are important for understanding and studying natural phenomena in the seas, oceans, in the mantle [6] and the atmosphere of the Earth. In addition, there is a wide range of processes and important applications where stratification with layered flows and segregation play a decisive role, for example, in the technological processes of obtaining new perfect materials from melts and solutions [7–8], environmental ecology, these are the tasks of storing and using liquid rocket fuel in tanks [4, 9]; problems of a boiling [10], problems cooling electronic devices and safety of nuclear energy [11], cleaning indoor air from pollution, smoke, as well as from finely dispersed liquid inclusions infected with viruses, in particular, COVID-19, etc. Convective processes of heat and mass stratification can have large- and micro-scale character with laminar and turbulent flows and their studies are determined by different goals. It is important to take into account the peculiarities of hydrodynamics and heat and mass transfer with heat and mass stratification not only in terrestrial conditions, but also in microgravity conditions [4].

These phenomena of stratification can be both negative and positive from the point of view of their use by a person. For example, when obtaining new materials, the macro-inhomogeneous distribution of impurities in the melt is a negative factor, since they tend to obtain single crystals with a uniform distribution of impurities across the ingot, as well as the difficulty of heat and mass transfer during stratified multi-vortex flow in volumes, seas and reservoirs, in the processes of obtaining materials, a boiling and cooling, etc. Density segregation and stratification in a liquid can also have a positive aspect, for example, when separating substances or obtaining eutectic materials with a certain structure.

Studies of stratification and formation of stratified convective structures are carried out using experimental, analytical and numerical methods. For slow flows, there are methods with analytical solutions and approximations, for example, [12–14], but experimental [15–18] and numerical methods [18-27, 31–34] are needed to reproduce the flow features under intense convection.

In this paper, examples of the formation of layered stratified stationary laminar flows caused only by thermal (concentrational) or thermo-concentration convection are considered [4, 5, 21–26].

The magnitude of the vertical temperature and concentration stratification varies non-linearly depending on the intensity of convection. This has been shown numerically and experimentally by different authors [4, 9, 17, 21–27].

In addition to gravitational convection, thermal and concentration-capillary convection can have an effect on heat and mass transfer [4, 28]. Marangoni capillary convection can be stable, unstable and have a threshold character [3, 4, 28, 29]. The paper points out that in conditions of weightlessness, capillary convection in a two-layer system has a metastable character and can change the direction of location the free surface and stratification [30].

## 2. Problem statement and mathematical model

The problem of thermal convection of an incompressible liquid in a horizontally layer with elongation $L/H$ from 1 to 12.71 (were $L$ – is length and $H$ – is height), laterally heated in the field of gravity (with acceleration of free fall **g**), is considered. At lateral heating, constant values of temperature $T_1$ and $T_2$ ($T_1 < T_2$) on the side walls are set; for velocities, non-slip conditions are set. The following boundary conditions are considered for horizontal walls: for velocity – the non-slip condition, , for temperature – the condition of thermal insulation $\partial T / \partial y |_{y=0, y=H} = 0$ is set. The scheme of the calculated geometry, boundary conditions and isotherms in layer for thermal conductivity case are shown in figure 1.

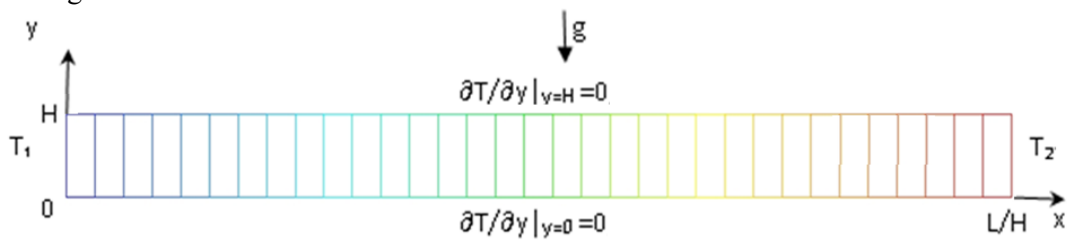

**Figure 1.** The calculated geometry, temperature boundary conditions and isotherms for heat conductivity regime.

The mathematical model is based on the numerical solution of a system of non-stationary planar 2D Navier-Stokes equations for natural convection of an incompressible liquid in the Boussinesq approximation [2, 26]:

$$\frac{\partial u}{\partial t} + u\frac{\partial u}{\partial x} + v\frac{\partial u}{\partial y} = -\frac{1}{\rho}\frac{\partial P}{\partial x} + \nu\left(\frac{\partial^2 u}{\partial x^2} + \frac{\partial^2 u}{\partial y^2}\right) + F_x, \quad (1)$$

$$\frac{\partial v}{\partial t} + u\frac{\partial v}{\partial x} + v\frac{\partial v}{\partial y} = -\frac{1}{\rho}\frac{\partial P}{\partial y} + \nu\left(\frac{\partial^2 v}{\partial x^2} + \frac{\partial^2 v}{\partial y^2}\right) + F_y, \quad (2)$$

$$\frac{\partial u}{\partial x} + \frac{\partial v}{\partial y} = 0, \quad (3)$$

$$\frac{\partial T}{\partial t} + u\frac{\partial T}{\partial x} + v\frac{\partial T}{\partial y} = a\left(\frac{\partial^2 T}{\partial x^2} + \frac{\partial^2 T}{\partial y^2}\right), \quad (4)$$

$$\frac{\partial C}{\partial t} + u\frac{\partial C}{\partial x} + v\frac{\partial C}{\partial y} = D\left(\frac{\partial^2 C}{\partial x^2} + \frac{\partial^2 C}{\partial y^2}\right) \quad (5)$$

where x, y – horizontal and vertical Cartesian coordinates; u, v – components of the velocity vector; t – time; T – temperature; C – concentration; $T_0$, $C_0$ – references value of temperature and concentration; P – pressure; $\rho$ – density; $F_x = -g_x\beta(T-T_0) - g_x\beta_C(C-C_0)$ and $F_y = -g_y\beta(T-T_0) - g_y\beta_C(C-C_0)$ – components of the vector of external forces; $\mathbf{g}(g_x, g_y)$ – vector of gravitational acceleration of the earth's free fall; $\beta$, $\beta_C$, $\nu$, $a$, $D$ – coefficients of temperature and concentration expansion of the liquid, kinematic viscosity, thermal conductivity and diffusion factor, respectively; in the future, we will use of dimensionless velocity and time (which was made dimensionless through the viscosity $\nu$ and height of the layer $H$).

The problem is characterized by dimensionless parameters: Rayleigh number $Ra = g\beta(T_2-T_1)H^3/\nu a$, concentrational Rayleigh number $Ra_c = g\beta_c(C_2-C_1)H^3/\nu D$, Prandtl number $Pr = a/\nu$, Schmidt number $Sc = \nu/D$, L/H aspect ratio. The Grashof number is equal $Gr = Ra/Pr$ and the concentrational Grashof number is equal $Gr_c = Ra_c/Sc$.

The results presented in this paper were obtained using different numerical methods: the finite-difference scalar method [31], the fully implicit matrix finite-difference approach [32], and the conservative control volume method [33]. The good accuracy of numerical results was confirmed by comparison with experimental data and comparison of numerical results obtained by different numerical models [2, 9, 17, 18, 32, 34].

In the case of modeling a two-layer system (paragraph 3.3), the VOF method was used with the solution of an additional transfer equation for the fraction function, and equations (1–4) were written for the "mixture" model with variable viscosity and thermal conductivity coefficients. The model is described in more detail in paper [30].

## 3. The results of numerical simulation

In this section of the paper, examples of the formation of layered stratified stationary laminar flows caused only by thermal (concentration) or thermo-concentration convection will be considered. In addition, it will be pointed out that in conditions of weightlessness, capillary convection in a two-layer system has a metastable character and can change the direction of the free surface and stratification.

### 3.1. Vertical stratification induced by gravitational convection

In figure 1 the isotherms presents in a liquid layer for thermal conductivity regime for without or very slow convection ($Ra < 10^3$). The solutions with convective single-vortex flows can be obtained analytically for given linear horizontal temperature distributions for small Rayleigh numbers, for example, see papers [12–14].

It should be noted that the approximation of the constantly of the temperature field (and the absence of vertical temperature stratification) in the fluid layer for the case of infinite temperature conductivity ($Pr = 0$) is not always true. This was showed and proved in paper [22].

*3.1.1. Vertical stratification induced by only thermal (or concentrational) convection*

In the case of slow convection ($Ra < 10^3$), there is no effect of convection on the temperature field and the value of vertical temperature segregation is zero or, depending on the value of the Prandtl number, the temperature field can be slightly changed by convection compared to the case of the thermal conductivity regime. For example, the simulation results in the form of isolines of the flow function ($\Psi$) and isotherms (T) of a convective flow with a single-vortex structure for $Ra = 2 \cdot 10^5$, $Pr = 5.8$, $L/H = 12.71$ in figure 2 are shown. Even with these parameters, thermal convection causes significant vertical temperature stratification, as can be seen in Fig. 2. In addition, near the vertical ends of the layer, the character of the temperature distribution is different than in the central part.

An approximate solution combining an analytical solution in the core with integral solutions at the ends of the layer was shown in [16]. But these solutions did not take into account the features of the multi-vortex flow near the vertical ends and the features of the flow in the core of the layer at large Rayleigh numbers. The features of the flow in long layers were shown in [21, 22].

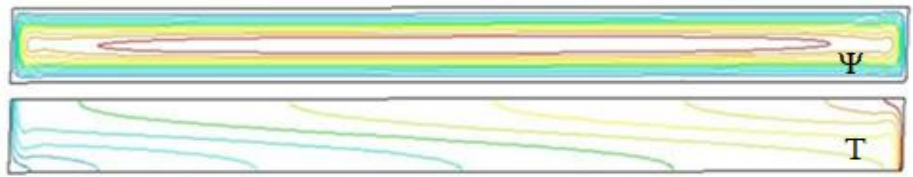

**Figure 2**. The isolines of the stream function ($\Psi$) and the isotherm (T) at $Ra = 2 \cdot 10^5$, $Pr = 5.8$, $L/H = 12.71$.

In figure 3 a development of convective flow in a horizontal layer (filled by water initially at the constant temperature $T_0 = (T_1 + T_2)/2$ after starting heating to the temperature $T_1$ of the right wall and cooling to the temperature $T_2$ of the left wall. Isolines of the stream function ($\Psi$) and isotherm (T) at different moments of dimensionless time: (a) – t=0.03, b) – 0.04, (b) – 0.05, d) – 0.64; (moments (a), (b), (c) – are nonstationary mode, (d) – is stationary mode). $Ra = 8 \cdot 10^7$, $Pr = 7.2$, $L/H = 6.94$ are shown. In figure 3 the isolines of the stream function and the isotherm at different points in time, and in figure 4 for this problem shows the dependencies of the maximum stream function and the averaged Nusselt number on the left side are shown.

In figure 3 on the initial moment of time (a), internal jets of the same intensity are formed on the vertical layer sides, which, accelerating and hitting to the horizontal walls (the lower one at the cold end and the upper one at the hot end) continue to move along the upper and lower walls towards each other (a). Then these two flows meet in the middle of the layer (b) and, practically without affecting each other, diverge and reach the opposite ends of the layer (c). Partially reflecting from the vertical walls and partially passing along the vertical "another" walls, these flows return to "their" end walls (where they were borne), but no longer strictly along the horizontal walls, but along a more complex trajectory due to the fact that part of the fluid was reflected from the ends. Thus, due to the viscosity, the fluid located in the central part of the layer is involved in the movement. In addition, new portions of liquid are also involved in the boundary layer on the vertical walls. Having reached "their" end walls. Having reached "their" end walls, the fluid flows return to the opposite ends, then back to "their", etc. Such a complex movement very delays the output of the convective flow to a stationary mode, making slow fluctuations of the main characteristics. Vertical stratification is established quickly, despite the fact that the velocity of temperature propagation is less, since the Prandtl number is greater than one ($Pr = 7.2$).

The fields of the stream function and temperature for the stationary mode are presented in figure 3 (d). The described moments are marked with markers (a–d) in figure 4 and it can be seen that the maximum stream function and the Nusselt number are sensitive to the specified flow moments. It should be noted that in the stationary mode, after the formation of vertical temperature stratification, secondary flows with countercurrents to the main flow are formed inside the layer (figure 3 (d)).

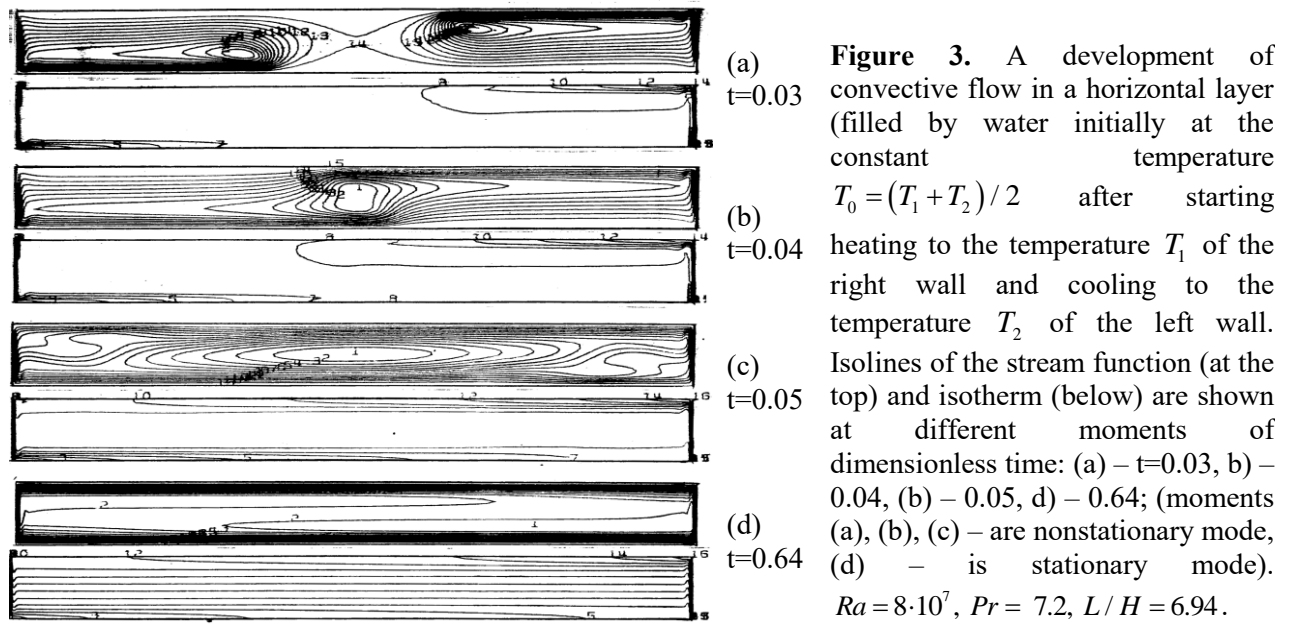

**Figure 3.** A development of convective flow in a horizontal layer (filled by water initially at the constant temperature $T_0 = (T_1 + T_2)/2$ after starting heating to the temperature $T_1$ of the right wall and cooling to the temperature $T_2$ of the left wall. Isolines of the stream function (at the top) and isotherm (below) are shown at different moments of dimensionless time: (a) – t=0.03, b) – 0.04, (b) – 0.05, d) – 0.64; (moments (a), (b), (c) – are nonstationary mode, (d) – is stationary mode). $Ra = 8 \cdot 10^7$, $Pr = 7.2$, $L/H = 6.94$.

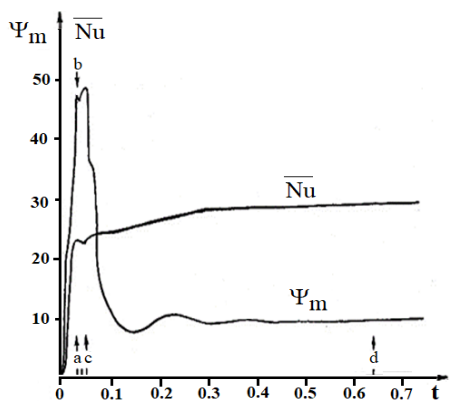

**Figure 4.** The dependencies of maximum stream function and the average Nusselt number (at the left side, x=0) on the time for $Ra = 8 \cdot 10^7$, $Pr = 7.2$, $L/H = 6.94$. The time points marked on the abscissa axis with arrows labeled (a), (b), (c) and (d) correspond to the same time moments shown in figure 3.

With an increase of the Rayleigh number up to $Ra = 10^9$, $Pr = 5.8\,109$ (Pr=5.8), convection practically remains stationary, the induced vertical stratification increases, secondary currents generate tertiary ones, as shown in figures 5 (b) and 6.

At high Rayleigh numbers (Ra>108), the flow is specific in that boundary layers generated on the horizontal and vertical walls and thin (vertically elongated) secondary vortices appear near the vertical ends. The temperature and dynamic boundary layers depend not only on the Rayleigh number, but also on the Prandtl number. In figure 5 isolines of the stream function (upper pictures) and the isotherm (bottom pictures) for different Rayleigh and Prandtl numbers for aspect ratio *L/H*=12.71: (a) – $Ra = 5.8 \cdot 10^8$, $Pr = 5.8$; (b) – $Ra = 10^9$, $Pr = 5.8$; (c) – $Ra = 10^9$, $Pr = 58$; (d) – $Ra = 1.7 \cdot 10^7$, $Pr = 0.1$ are shown. The results are presented in figure 5 show that the width of the boundary layers and the degree of vertical stratification vary but secondary layered flows remain in both cases: for large ((c) – $Ra = 10^9$, $Pr = 58$) and for small Prandtl numbers, ((d) – $Ra = 1.7 \cdot 10^7$, $Pr = 0.1$). More details about the influence of Rayleigh and Prandtl numbers on convection in horizontal layers are described in the paper [22].

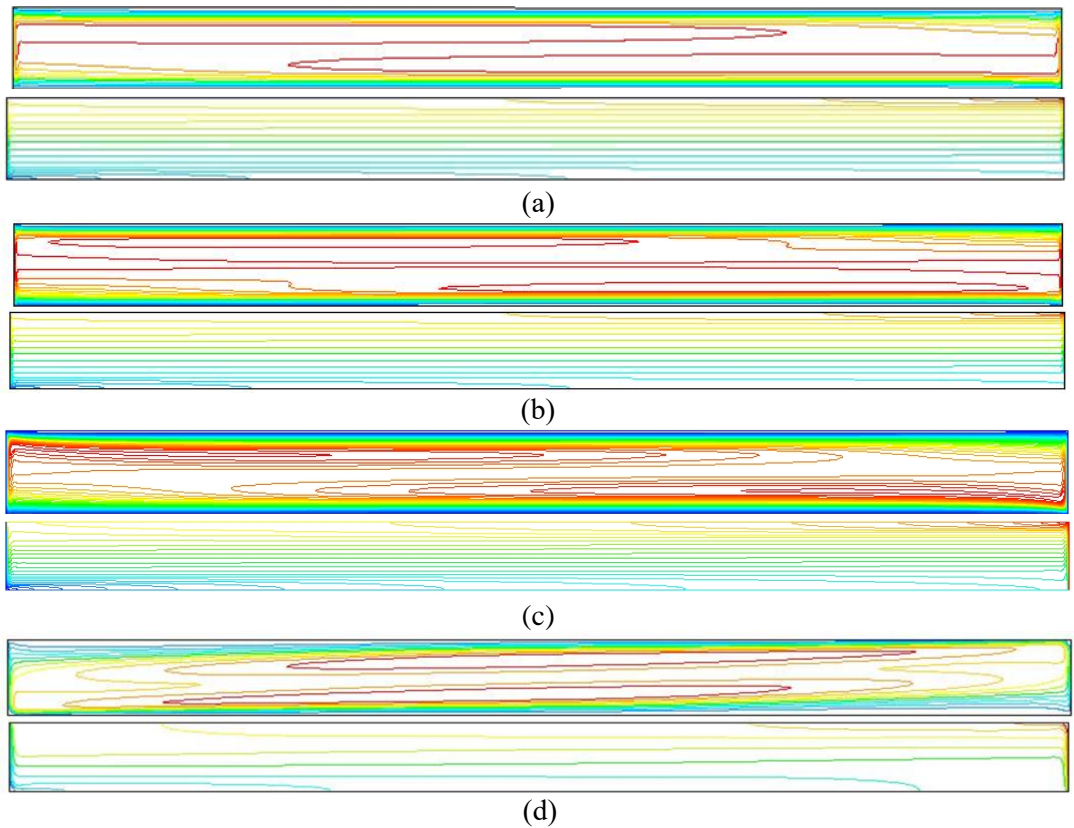

**Figure 5.** The isolines of the stream function (upper pictures) and the isotherm (bottom pictures) at *L/H*=12.71: (a) – Ra=5.8 10⁸, Pr= 5.8; (b) – $Ra = 10^9$, $Pr = 5.8$; (c) – $Ra = 10^9$, $Pr = 58$; (d) – $Ra = 1.7 \cdot 10^7$, $Pr = 0.1$.

The profiles of velocities for layer with aspect ratio *L/H*=12.71: (a) - the horizontal component dimensionless velocity u in the middle vertical cross section (x=6.355) for various parameters: line $1 - Ra = 2 \cdot 10^5$, $Pr = 5.8$; $2 - Ra = 1.2 \cdot 10^8$, $Pr = 5.8$; $3 - Ra = 10^9$, $Pr = 5.8$; $4 - Ra = 1.2 \cdot 10^8$, $Pr = 58$; (b) – profiles of the vertical dimensionless velocity component in the average horizontal section (y=0.5) near the left end of the layer at, obtained by $Ra = 10^9$, $Pr = 5.8$ different numerical methods (the line is the finite difference method, the points are the control volume method) are present in figure 6. These results show the influence of Rayleigh and Prandtl numbers on the velocities of the main and secondary flows. In figure 6 (b) the presence of a lifting velocity component near the left cooled vertical wall, that indicates the presence of a secondary vortex are shown.

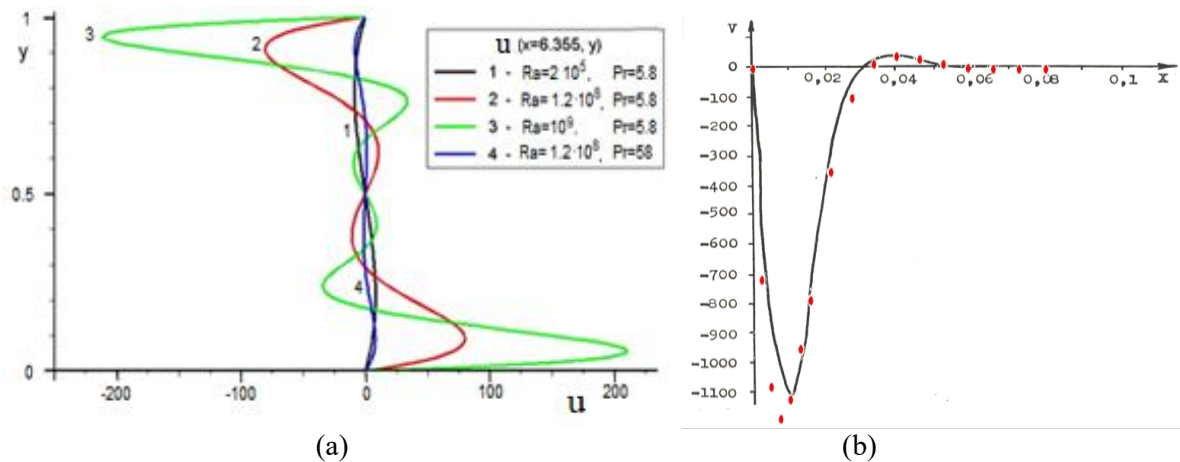

**Figure 6.** The profiles of velocities for layer with aspect ratio $L/H=12.71$:
(a) - the horizontal component velocity u in the middle vertical cross section (x=6.355) for various parameters: line $1 - Ra = 2 \cdot 10^5$, $Pr = 5.8$; $2 - Ra = 1.2 \cdot 10^8$, $Pr = 5.8$; $3 - Ra = 10^9$, $Pr = 5.8$; $4 - Ra = 1.2 \cdot 10^8$, $Pr = 58$; (b) - profiles of the vertical velocity component in the average horizontal section (y=0.5) near the left end of the layer at, obtained by $Ra = 10^9$, $Pr = 5.8$ different numerical methods: the line is the finite difference method, the points are the control volume method.

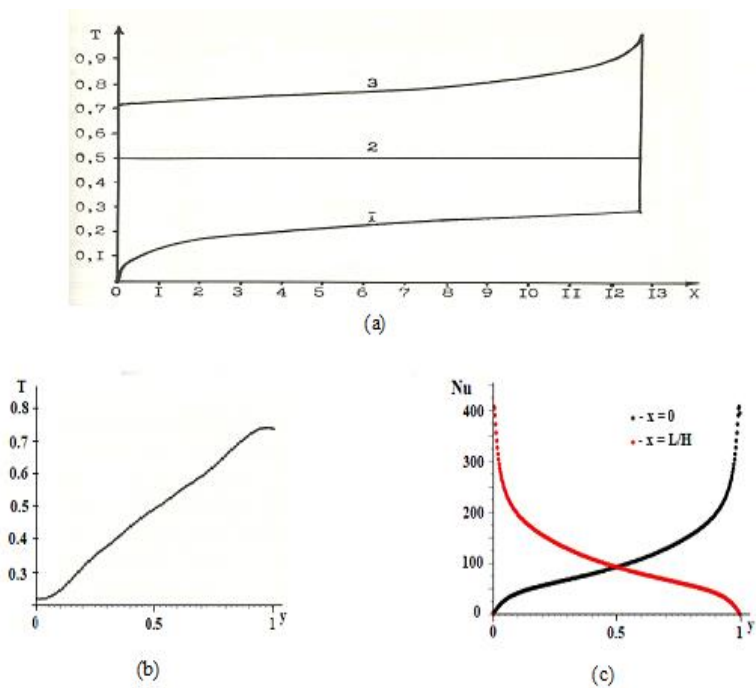

**Figure 7.** Temperature profiles and dimensionless heat fluxes; (a) – horizontal temperature profiles: line 1 – on the bottom wall (y =0), 2 – along horizontal middle section (y=0.5), 3 – on the upper wall (y=1); (b) – temperature profile along vertical middle section (x=6.355); (c) – distribution of the Nusselt number at left and right ends of the layer. $Ra = 10^9$, $Pr = 5.8, L/H = 12.71$.

In figure 7 temperature profiles and dimensionless heat fluxes for $Ra = 10^9$, $Pr = 5.8, L/H = 12.71$ are present. Temperature profiles along the layer show that in the middle section the temperature is constant almost everywhere except at the ends, and change slightly along the upper and lower walls (figure 6 (a)). The results showed that approximately 80 of the temperature change occurs at a distance of 0.02 L/H at the end wall. The vertical temperature profile (figure 6 (b)) illustrates the vertical temperature stratification and show that secondary currents do not greatly affect it. The heat flux at the left and at right vertical walls are symmetrical, these decrease and increase, respectively (figure 6 (c)),

*3.1.2. Vertical stratification induced by thermo-concentration convection*
In figure 8 shows the vertical temperature and concentration stratifications of the liquid in stationary modes with lateral heating and a vertical concentration gradient (concentrations are given: $C_2=1$ – on the upper wall, $C_1=0$ – on the lower wall, heavy at the bottom). Figure 1 shows the results of thermo-concentration convection with lateral heating and a vertical concentration gradient at the Grashof numbers $Gr = 10^6$, $Gr_c = 4 \cdot 10^6$, Prandtl Pr=10. In figure 8 on the left: isolines of concentration (C), temperature (T) and current functions ($\Psi$) for three Schmidt numbers (on columns): (a) – $Sc = \nu/D = 1$, (b) – $Sc = 10$, (c) – $Sc = 100$) are shown. Figure 8 on the right temperature profiles (T – solid lines) and concentration profiles (C – dotted lines) in the middle vertical section also for three Schmidt numbers ($Sc = 1, 10, 100$) are shown.

The results show in figure 8 for steady-state stationary mode. The times of establishing the flow velocity, temperature and concentration fields are different, so the calculation was carried out before all three values were stationary. It can be seen that layered structures are formed in all three cases, but their structure and layering strongly depends on the Schmidt number for given Grashof numbers: $Gr = 10^6$, $Gr_c = 4 \cdot 10^6$.

Thus, it is possible to control the created stratification and segregation of temperature or concentration by changing the defining dimensionless parameters (Schmidt number or Rayleigh number)

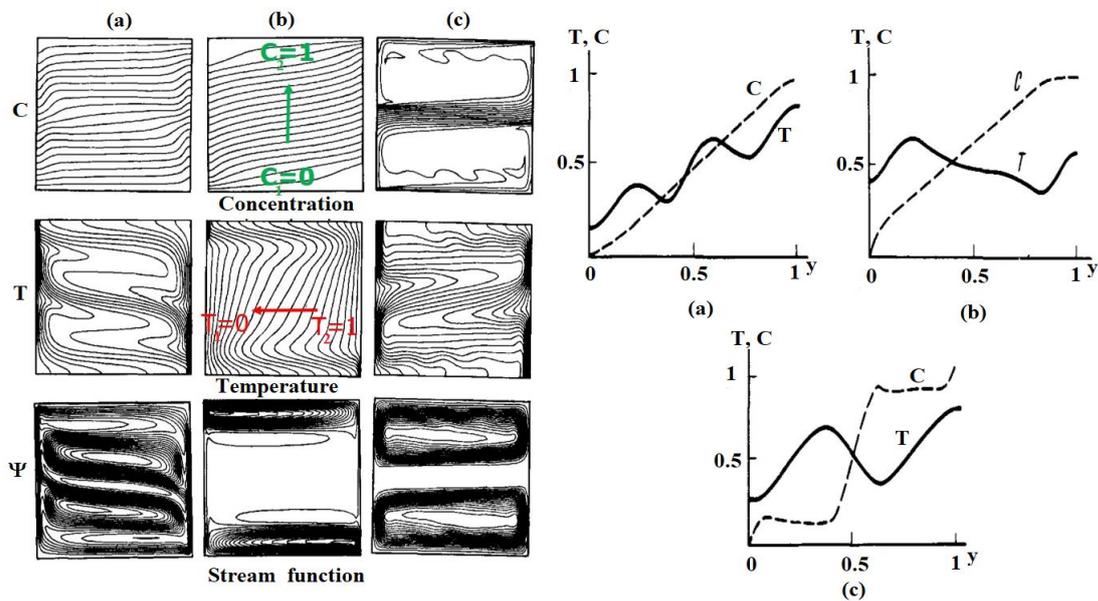

**Figure 8**. Thermo-concentration convection with lateral heating and vertical mass flux at Grashof numbers $Gr = 10^6$, $Gr_c = 4 \cdot 10^6$ and $Pr = 10$, $L/H = 1$ for various Schmidt numbers: (a) – $Sc = 1$, (b) – $Sc = 10$, (c) – $Sc = 100$; on the left of figure – isolines of concentration (C), temperature (T) and stream function ($\Psi$) are show; on the right – profiles of temperature and concentration in the middle vertical section vertical cross section for three Schmidt number.

Previously conducted experiments on Earth and in Space under the EURECA program (1992,1993) on the crystallization of calcium phosphates from solutions showed that the size of hydroxyapatite crystals grown in zero gravity is ten to a hundred times larger than their terrestrial counterparts.

The results of numerical simulation of these experiments on the crystallization of calcium phosphates in a thermostat have shown possible convective mechanisms and convective structures of the transfer of reaction components on Earth and in microgravity conditions [5]. It was shown that in terrestrial conditions it is possible to form layered convective flow structures in reactor similar to those shown in figure 8, which can lock and reduce the transport of components for a chemical reaction, which could explain the unusual experimental results obtained under the EURECA program.

*3.2. The dependences of the temperature and concentration stratification on the determining dimensionless parameters*

The magnitude of the temperature and concentration stratification vary non-linearly depending on the determining dimensionless parameters, in particular on the intensity of convection [5, 9, 23]. It was found that the dependence of the vertical stratification (stratification) on the Rayleigh number has a maximum. This effect of the maximum concentration stratification was manifested in the space experiment MA-150 "Universal Furnace" for the unidirectional crystallization of germanium, performed under the "Soyuz-Apollo" program.

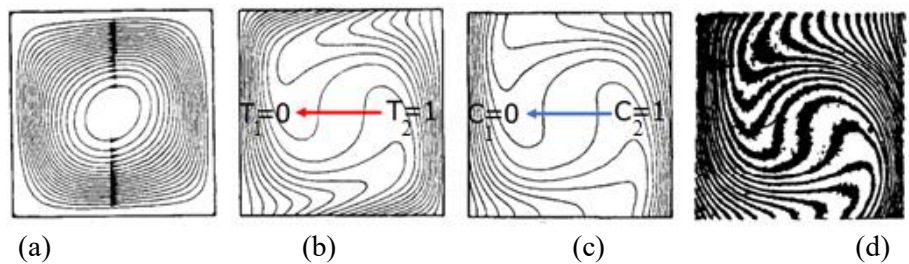

(a)      (b)      (c)      (d)

**Figure 9.** Thermo-concentration convection in a square area with lateral heating and a horizontal concentration gradient. (a) – isolines of the stream function, (b) – isotherms, (c) – lines of equal concentrations and (d) - interferogram of the density field [17].
$Ra = 1.5 \cdot 10^5$, $Ra_c = 1.2 \cdot 10^5$, $Pr = 0.75$, $Sc = 0.65$.

The results of this experiment showed that the radial inhomogeneity of the alloying impurity in the sample obtained in Space was several times greater than in a similar sample obtained in the terrestrial experiment. To verify this result, special hydrodynamic experiments [17] and verification numerical calculations were carried out. These experiments were carried out on gases in a cubic cavity heated from the side and with the impurity flux from the side also. The calculations were carried out for the square cross section of the experimental cubic cavity (figure. 9). In figures 9–10 comparisons of the simulation results of concentration stratification with experimental data [17] in stationary mode with the same experimental and modelling dimensionless parameters are shown for Rayleigh number $Ra = 1.5 \cdot 10^5$, Rayleigh concentration number $Ra_c = 1.2 \cdot 10^5$, Prandtl number $Pr = 0.75$, Schmidt number $Sc = 0.65$. In figure 9 the isolines of the current function, isotherms, lines of equal concentrations and interferogram of the density field are presented [17].

In figure 10 the dependence of the concentration derivative on the vertical coordinate calculated in the center of the square area on the Rayleigh number for thermal, concentration and thermo-concentration convection, for three cases: (a) – thermal convection, solid line 1 – calculation, dotted line 2 – experiment [17], $Ra_c = 0$, $Pr = Sc = 0.75$; (b) – concentration convection, different icons - experimental data for different gas mixtures [17], solid line 1 – calculation at $Ra = 0$, $Sc = 1$; (c) – thermo-concentration convection, points 2 – experiment [17], solid line 1 – calculation at $Ra_c = 1.2 \cdot 10^5$, $Pr = 0.75$, $Sc = 0.65$ are shown It can be seen from figure 10 that the experimental and numerical results showed good agreement and confirmed the existence of a concentration stratification maximum

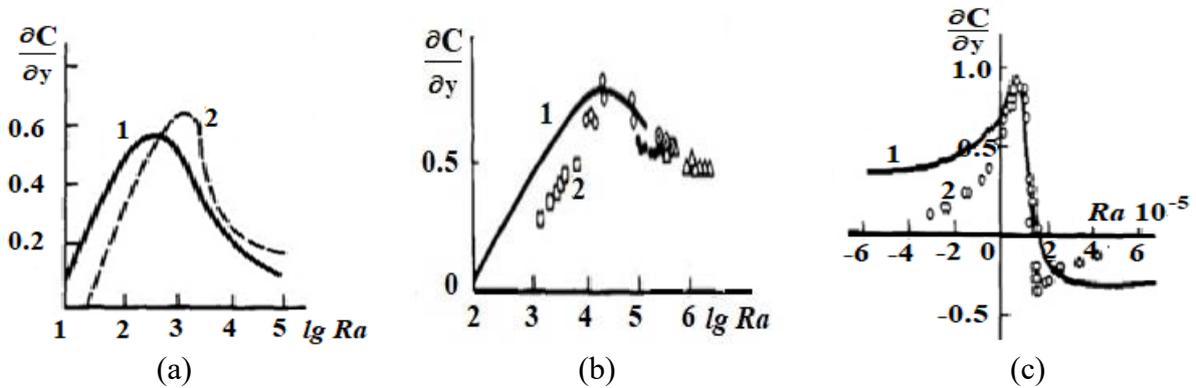

**Figure 10.** The dependencies of the derivative of the concentration along the vertical coordinate $\partial C/\partial y$ in the center of the square region on the Rayleigh number for thermal – (a), concentration – (b) and thermo-concentration – (c) convections.

*3.3. Changing the direction of density stratification of a two-layer "water-air" system under zero gravity.*

As is known, in zero gravity conditions, capillary convection can occur in liquid volumes with a free boundary, caused by a temperature gradient or an impurity concentration [4, 28]. The capillary flow can be stable or unstable, stationary or unsteady, and in some cases a hydrodynamic system with a thermocapillary convective flow can be metastable. Marangoni convection can have a threshold effect. The reasons for the threshold effect may be: the directions of the heat flows or (and) concentrations, the rheological properties of the liquid, external conditions and contact properties, if solids are present [29]. The study of the behavior of a free surface in two-layer liquid systems is important not only for fundamental studies of hydrodynamics, but also for solving many applied problems, for example, for aviation and astronautics, healthcare, chemical industry, technologies for obtaining materials and medicines, etc. [4, 28]. In this section of the article, we will consider an example of metastable instability of the interface of a two-layer water-air system in weightlessness in the presence of thermocapillary convection.

Consider the case of thermocapillary convection in a square region with frictionless boundaries half filled with water and air. The problem statement is described in [30] and shown in the figure.11 (a). Constant temperatures $T_1$ and $T_2$ ($T_1 < T_2$) are set on the side walls. Thermal insulation conditions are set on the horizontal walls. Due to the surface tension at the interface, capillary convection occurs. Due to intense capillary convection, the interface is slightly curved and weakly oscillates in time. Over time, the air and water warm up. Isotherms after warming up acquire mainly a conventionally vertical direction (figure.11 (b)). By this time, the conditionally horizontal location of the interface becomes metastable unstable. Then this position of the interface, with a weakly oscillating change, can remain for quite a long time. But with a small short-term disturbance (vibrations, accelerated rotation of the entire area or changes in the intensity of convection, etc.), the free interface can turn sharply by 90 degrees and remain in this position, oscillating near the vertical location [30]. The free surface assumes an energetically advantageous location - a predominantly vertical position (adjusting along the isotherms along which the intensity of thermocapillary convection is minimized). Thus, the direction of density stratification in a two-layer system changes by 90 degrees.

The simulation results are shown in figure 12 ($\text{Ma} = 10^6$, Ra=0) in the form of colored isolines of the values of the volume fraction of the liquid (a) and isotherms (b) averaged over time in the interval from t = 0 to t =16 seconds. The interface performs weakly damped oscillations near the vertical. It should be noted that if gravity is "turned on", the interface returns to its original horizontal position.

This effect is absent in terrestrial conditions, since gravitational convection has a damping effect on the position of the free surface [30].

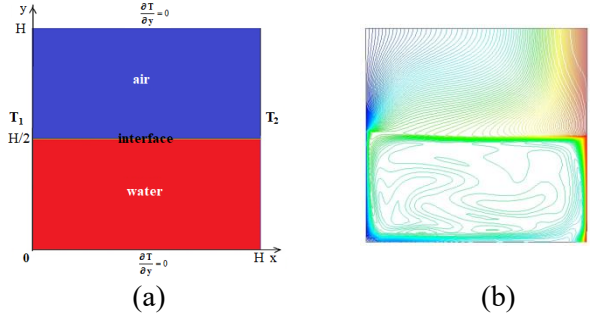

**Figure 11.** Location of the air-water interface (a) and the isotherm (b), for time *t*=5 sec, $Ma = 10^6$, $Ra = 0$.

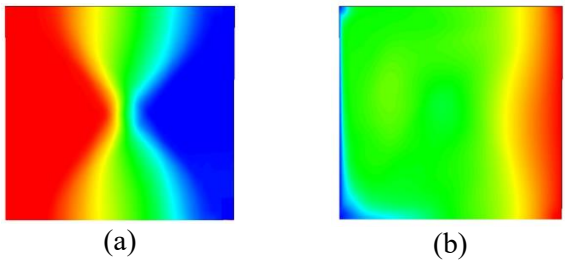

**Figure 12.** Contours of the volume fraction of water (a) and the isotherms (b) averaged over time across the range from *t*=0 to *t*=15.86 sec ($Ma = 10^6$, $Ra = 0$).

## 4. Conclusions

In horizontal layer filled with incompressible liquid with lateral heated due to thermal convection, the initially homogeneous temperature field redistribute into a field with strong stable vertical stratification. In such layers of fluid with stable stratification on density (induced only by thermal (or only concentration) convection) arise horizontal layered convective structures with counter streams flow to the main flow was shown.

Analysis of the results of the considered problems has shown that even with laminar stationary convection, nonlinear features of the fluid flow can have a significant impact on the vertical temperature and concentration stratification. It is shown that the formation of stratified structures can occur during laminar convection, both in homogeneous and inhomogeneous media. Thus, firstly, the formation of stratified flows can be carried out in a homogeneous liquid (or in an isothermal, but inhomogeneous fluid), and the simultaneous presence of double diffuse convection mechanisms can only aggravate or weaken stratified flows and liquid stratifications. Secondly, the reason for the formation of stratified flows is not small-scale turbulent or oscillatory flows, but large-scale convective flows that create vertical density stratification.

The magnitude of the vertical stratification depends on the properties of the liquid and depends nonmonotonically on the intensity of the convective flow, for example, with an increase in the Rayleigh number, the magnitude of the average vertical stratification increases and then decreases, so that a maximum is observed.

It is shown that under conditions of weightlessness, the changing in the direction of stratification of the density of the two-layer air-liquid system (with its lateral heating and the presence of capillary convection) has a metastable character. Under certain perturbations, the position of the free boundary can change sharply by 90 degrees, moving to a stable location. Gravity eliminates this effect.

**Acknowledgements**
*The study was supported by the Government program (contract # AAAA-A20-120011690131-7) and was funded by RFBR, project number 20-04-60128.*